\documentclass[aps,pra,twocolumn,groupedaddress]{revtex4} %submit to arXiv

\usepackage{graphicx}
\usepackage{bm}
\usepackage{array}
\usepackage{amsmath}
\usepackage{subfigure}
\usepackage{ulem}

\begin{document}

%\preprint{}

\title{Controlling Condensate Collapse and Expansion with an Optical Feshbach Resonance}



\author{Mi Yan, B. J. DeSalvo, B. Ramachandhran, H. Pu, and T. C. Killian}

%\thanks{}
\affiliation{Rice University, Department of Physics and Astronomy, Houston,
Texas, 77251}

%\author{P. Pellegrini, and R. C\^ot\'e}

%\thanks{}
%\affiliation{$^2$Department of Physics, U-3046, University of
%Connecticut, Storrs, CT, 06269-3046}


\date{\today}

\begin{abstract}
We demonstrate control of the collapse and expansion of an $^{88}$Sr Bose-Einstein condensate using an optical Feshbach resonance (OFR) near the $^{1}S_{0}$-$^{3}P_{1}$ intercombination transition at 689\,nm. Significant changes in dynamics are caused by modifications of scattering length by up to $\pm 10\,a_{\textrm{bg}}$, where the background scattering length of $^{88}$Sr is  $a_{{\textrm{bg}}} = -2\,a_0$ ($1\, a_0=0.053\,$nm).
%can be achieved within one hundred molecular  linewidths detuning from a photoassociative  line with small atom losses.
Changes in scattering length are monitored through changes in the size of the condensate after a time-of-flight measurement. Because the background scattering length is close to zero, blue detuning of the OFR laser with respect to a photoassociative resonance leads to increased interaction energy and a faster condensate expansion, while red detuning triggers a collapse of the condensate. The results are modeled with the time-dependent non-linear Gross-Pitaevskii equation. %This allows us to extract parameters that describe the OFR and collisions in a light field at large detuning from atomic resonance.

\end{abstract}

%\pacs{32.80.Pj}
%\keywords{}

\maketitle

%\section{Introduction\label{introduction}}
%\subsection{}
%\subsubsection{}

% If in two-column mode, this environment will change to single-column
% format so that long equations can be displayed. Use
%\begin{widetext}
%\end{widetext}

The ability to tune interactions in ultracold atomic gases  makes these systems
ideal for exploring many-body physics \cite{bdz08} and has enabled some of the
most important recent advances in atomic physics, such as investigation of the Bose-Einstein condensate
(BEC)-Bardeen-Cooper-Schrieffer   crossover regime \cite{bdz08} and creation of quantum degenerate
molecules \cite{grj03,jba03}. Magnetic Feshbach resonances \cite{cgj10}, which are the standard tool for
changing atomic interactions, have proven incredibly powerful, but they are
also limited because the methods for creating magnetic fields preclude high-frequency spatial and temporal modulation.
Also, in atoms with non-degenerate ground states, such as alkaline-earth-metal atoms, magnetic Feshbach resonances do not exist.

These limitations can be overcome by using an optical Feshbach resonance (OFR), which tunes interatomic interactions by coupling a colliding atom pair to a
bound molecular level of an excited state potential with a laser tuned near a
photoassociative resonance \cite{fks96}. Optical Feshbach resonances may open
new avenues of research in nonlinear matter waves \cite{sue03,rmp05,kmt11} and
quantum fluids \cite{fwg89,qzh11,chi11},
and could be very valuable for experiments with fermionic alkaline-earth atoms \cite{dym10,tsg10} in lattices \cite{sit11}, which possess SU(N) symmetry with large N and have attracted great attention lately because of novel  thermodynamics \cite{hgh12,chw12,tys12} and predictions of frustrated magnetism and topological ground states \cite{whz03, chu09, hho04, hgr09}.
Here we present the control of collapse and expansion of an $^{88}$Sr BEC with an OFR near the
$^{1}S_{0}$-$^{3}P_{1}$ intercombination transition at 689 nm.

Early experiments on OFRs \cite{fjl00,ttw04,ttw05} used strong dipole-allowed
transitions in alkali-metal atoms to alter atomic collision properties, but substantial change in the atom-atom scattering length was accompanied by rapid atom losses.
Tuning of interactions in alkali-metal atoms, but with smaller atom loss, was recently obtained with a  magnetic Feshbach resonance using an AC Stark shift of the closed channel to modify the position of the resonance \cite{blv09,Bauer2009}.
Recently, a multiple-laser optical method was proposed for wider modulation of the interaction strength near a magnetic Feshbach resonance \cite{wth12}.
Unfortunately, none of these hybrid variations are feasible for atoms lacking magnetic Feshbach resonances.

Ciurylo \textit{et al.} \cite{ctj05,ctj06} predicted that an OFR induced by a
laser tuned near a weakly allowed transition should  tune the scattering length
with significantly less induced losses. This can be done with divalent atoms,
such as strontium and ytterbium, by exciting near an intercombination
transition from the singlet ground state to a metastable triplet level. The
improved OFR properties result from the long lifetime of the excited molecular
state and relatively large overlap integral between excited molecular and
ground collisional wave functions. Intercombination-transition OFRs have been
used to modify the photoassociation (PA) spectrum in a thermal gas of Yb
\cite{ekk08}, modulate the mean field energy in a Yb BEC in an OFR-laser
standing wave \cite{yts10}, and modify thermalization and loss rates in a
thermal gas of $^{88}$Sr \cite{bnb11}. In the OFR work with an Yb BEC
\cite{yts10}, small detunings from a molecular resonance were used
($\left|\Delta\right| < 10\,\Gamma_{\textrm{mol}}$, where
$\Gamma_{\textrm{mol}}$ is the natural decay rate of the excited molecular
level), which led to short sample lifetimes on the order of microseconds.
Longer exposure times and detunings $\left|\Delta\right| < 50\,\Gamma_{\textrm{mol}}$ were used in thermal Sr gases \cite{bnb11}, but at
much lower atomic density than typically found in a degenerate sample.

There is great interest in  intercombination-line OFRs at much larger detuning  in quantum degenerate gases
 of divalent atoms \cite{mmy09,sth09,mmy10,dym10},
 with the goal of modifying the scattering length and still maintaining  sample lifetimes  on the order of dynamical timescales of quantum fluids \cite{qzh11,chi11}. Here we use an OFR to control  collapse and expansion of an $^{88}$Sr condensate during time-of-flight measurements.
$^{88}$Sr has an $s$-wave background scattering length of $a_{{\textrm{bg}}} = -2\,a_0$ \cite{mmp08,skt10}, which allows convenient modification of the scattering length  either positive or more negative. Large relative change in scattering length $a_{\textrm{opt}}/a_{{\textrm{bg}}}=\pm10$ is demonstrated, with the loss-rate constant $K_{\textrm{in}} \sim 10^{-12}$ cm$^3$/s comparable to Ref.\,\cite{Bauer2009}.
We explore $\left|\Delta\right|$ as large as $667\,\Gamma_{\textrm{mol}}$, and  obtain sample lifetimes of milliseconds during application of the OFR beam.

According to the isolated resonance model \cite{ctj05,ctj06}, a laser of wavelength $\lambda$  detuned by $\Delta$ from  a photoassociative transition to an excited molecular state $|n\rangle$ modifies the atomic scattering length according to
$a=a_{{\textrm{bg}}}+a_{\textrm{opt}}$ and induces
two-body inelastic collisional losses
described by the loss rate constant $K_{\textrm{in}}$, where
\begin{eqnarray}\label{OFRFormulas}
a_{\textrm{opt}}&=& \frac{\ell_{\textrm{opt}}\Gamma_{\textrm{mol}}\Delta}{\Delta^2+\frac{(\eta\Gamma_{\textrm{mol}})^2}{4}};
\nonumber\\
K_{\textrm{in}}&=&
\frac{2\pi\hbar}{\mu} \frac{\ell_{\textrm{opt}}\eta\Gamma_{\textrm{mol}}^2}{\Delta^2+\frac{(\eta\Gamma_{\textrm{mol}}+\Gamma_{\textrm{stim}})^2}{4}}.
\end{eqnarray}
$K_{\textrm{in}}$ is defined such that it contributes to the
evolution of density $n$ as $\dot{n}=-K_{\textrm{in}} n^{2}$ for a BEC. The optical length $\ell_{\textrm{opt}}$, which characterizes the strength of the OFR, is defined as
\begin{equation}\label{loptical}
\ell_{\textrm{opt}}=\frac{\lambda^3 |\langle n | \varepsilon_r \rangle |^2 I}{16 \pi c k_{r}},
\end{equation}
where $c$ is the speed of light, $I$ is the intensity of the OFR beam, and $k_r$ is the wavenumber for colliding atoms, given by $k_r=\sqrt{21/8}/(2R_{\textrm{TF}})$  for a BEC with Thomas-Fermi radius $R_{\textrm{TF}}$, and $k_r=\sqrt{2\mu\varepsilon_{r}}/\hbar$ for a thermal gas, where $\mu=m/2$ is the reduced mass for the atomic mass $m$, $\varepsilon_r$ is the kinetic energy of the colliding atom pair, and $\hbar$ is the reduced Planck constant.
$|\langle n | \varepsilon_r \rangle |^2$ is the Franck-Condon
factor per unit energy for the free-bound PA transition. Because $|\langle n | \varepsilon_r \rangle |^2\sim k_r$ in the ultracold regime \cite{bju99}, following the Wigner threshold law, $\ell_{\textrm{opt}}$ is independent of the collision energy.
$\Gamma_{\textrm{mol}} =2\pi \times 15$\,kHz is the natural linewidth of the excited molecular level, and $\Gamma_{\textrm{stim}} = 2k_r \ell_{\textrm{opt}} \Gamma_{\textrm{mol}}$ is the laser-stimulated linewidth.  The parameter $\eta > 1$   accounts for enhanced molecular losses, as observed in previous OFR experiments \cite{ttw04,bnb11}.

 As shown through coupled channels calculations \cite{bnb11}, the isolated-resonance-model expressions (Eq.\,\ref{OFRFormulas}) break down at large detunings from photoassociative resonance. The induced scattering length $a_{\textrm{opt}}$ crosses zero between resonances. Outside approximately 100 linewidths from photoassociative resonance, the two-body loss is expected to make a transition to a broad background value that varies as $1/\delta^2$, where $\delta$ is $2\pi$ times the detuning from atomic resonance \cite{bnb11}. A rigorous theoretical description for loss in this regime is lacking, but the underlying mechanism  is  collisions involving a ground state atom and an atom excited  in the wings of the atomic line. In the regime where molecular levels are unresolved, such as in light-assisted collisions in a magneto-optical trap, this loss is often described with the classical Gallagher-Pritchard model \cite{weiner}. In a coupled channels description, the background loss rate is sensitive to a cutoff atom-atom distance inside of which radiative loss is turned on, which is introduced as an \textit{ad hoc} parameter \cite{julienne}. Our measurements could provide some experimental input to determine this cutoff distance.
 We find the isolated-resonance-model expressions (Eq.\,\ref{OFRFormulas}) useful for describing our measurements with the modification that the total loss rate constant is given by $K_{\textrm{total}}=K_{\textrm{in}}+K_\textrm{b}$, where the background loss is described phenomenologically in our regime as
 $K_\textrm{b}=K_0[\Gamma_{\mathrm{mol}}/(2\delta)]^2$.
 %\begin{equation}\label{eq:backgroundloss}
%  K_b=K_0(\delta_0/\delta)^2.
% \end{equation}

 %The parameter $\eta > 1$  in Eq.\ \ref{OFRFormulas} accounts for extra molecular losses observed in previous OFR experiments \cite{ttw04,bnb11}. These experiments probed the core  of the photoassociative line ($|\Delta|<50 \Gamma_{\textrm{mol}}$), and found that $\eta\sim 3$. Our experiments, however, probe larger detuning in which the loss is transitioning from being dominated by the core $K_{in}$ to being dominated by the background $K_b$. We find that a much larger value of $\eta$ is needed to describe observations in this regime. Alternatively, one could interpret this as meaning that the far wings of spectrum of photoassociative loss are not well described by a Lorentzian.
% %For the laser intensities and detunings in our study, $a_{\textrm{opt}}$ is not sensitive to $\eta$.

To probe the change in scattering length and loss, we monitor  expansion of an
$^{88}$Sr BEC after release from the optical dipole trap (ODT) with time-of-flight absorption
imaging using the $^1S_0$-$^1P_1$ transition. Details of the formation of an
$^{88}$Sr BEC are given in Ref.\,\cite{mmy10}. We create condensates with about
 7000 atoms, size $\sigma_0=0.8$\,$\mu$m, and peak density $n_0=1\times 10^{15}\,\mathrm{cm}^{-3}$. About 10\% of the trapped atoms are in the condensate and this represents
about 95\%
of the critical number for collapse with the background
scattering length of $^{88}$Sr for our ODT, which is
close to spherically symmetric with the geometric mean of the trap oscillation
frequency $\overline{\omega} = 2\pi\times(60 \pm 5)\,$Hz \cite{ycm11}.  The 689\,nm OFR laser
beam is tuned near the photoassociative transition to the second least bound
vibrational level on the
 $^1S_0$+$^3P_1$ molecular potential, which has the binding energy of $h\times 24$\,MHz \cite{zbl06}.

The OFR laser, with a beam waist of $725\,\mu$m, is applied to the condensate
20\,$\mu$s before extinguishing the ODT and left on for a variable time $\tau$
during expansion. The exposure time in the ODT is short
enough that the initial density distribution of the condensate reflects the ODT
potential and the background scattering length, while the expansion dynamics
is sensitive to the interaction energy determined by
$a=a_{{\textrm{bg}}}+a_{\textrm{opt}}$.

%\begin{figure}
%\includegraphics[clip=true,keepaspectratio=true,width=3.4in]{figure1.eps}\\
%\caption{(color online) Timing diagram (partial) for the optical Feshbach resonance experiment. Here 0 ms is at the moment when the ODT beams are extinguished, and the exposure time of the OFR beam $\tau$ is on the order of several ms.
% \label{OFRtiming} }
%\end{figure}

\begin{figure}
\includegraphics[trim = 10mm 60mm 0mm 0mm, clip=true,keepaspectratio=true,height=1.3in]{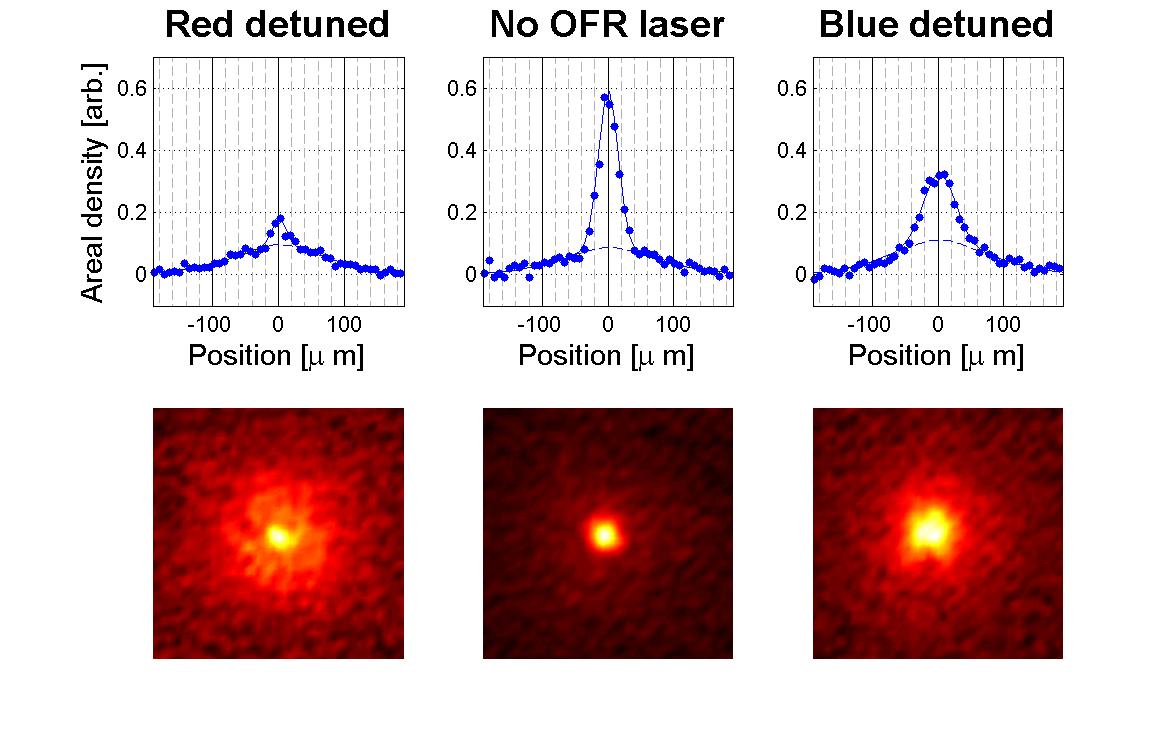}\\ %clip=true,keepaspectratio=true,height=2.3in]{figure2.eps}\\
\caption{(color online) Line profiles through absorption images showing OFR-induced variation of BEC expansion. Data correspond to no OFR laser and an OFR laser blue and red detuned by 0.5\,MHz  with respect to the -24\,MHz PA line \cite{zbl06} applied for $\tau=1.2$\,ms. Expansion times are 35\,ms. Fits are a Bose distribution for the thermal atoms (- -) and a Gaussian density distribution for the BEC.
\label{feshbachonoff} }
\end{figure}

Figure\,\ref{feshbachonoff} shows 1D slices through absorption images of atoms after a 35\,ms time-of-flight with and without application of the OFR laser.
Absorption images measure the areal density, which is
fit with a bimodal function including a Bose distribution for the thermal atoms and a narrow gaussian density distribution for the BEC, $n(r)=\frac{N_{0}}{2\pi\sigma^2 } \exp\left[-\frac{r^2}{2 \sigma^2}\right]$, to determine the number of atoms in the BEC $N_{0}$ and BEC size $\sigma$. (Quoted sizes reflect
correction for  imaging system resolution, which is modeled by a point spread function  $L(r)=\frac{1}{2\pi s^2 } \exp\left[-\frac{r^2}{2 s^2}\right]$ with $s= 5\pm 1$\,$\mu$m.)
The condensate size after a long time of flight is a good probe of interactions because of the sensitivity to the initial interaction energy.

To obtain a qualitative understanding of the data, one can
calculate the total energy immediately after the trap is extinguished using the
condensate energy functional \cite{pmc97,dgp99} assuming a gaussian density for the BEC in the ODT with initial size $\sigma_0$.
When atom losses are negligible, this  energy can be equated to the total kinetic energy when the condensate has expanded to a low density to give,
\begin{eqnarray}\label{OFR simple model}
N_{0} \frac{3}{2}m\sigma^{2}_{v}&=&
N_{0} \frac{3}{8}\frac{\hbar^{2}}{m\sigma_0^2}+ N_{0}^2 \frac{g}{2(4\pi)^{3/2}\sigma_0^3}.
\end{eqnarray}
The first and second terms on the right-hand side are the kinetic energy and interaction energy in the trap before release, respectively, for $g=4\pi \hbar^2 a/m$. $\sigma_{v}$ is the rms velocity, which can be related to the BEC size after a long expansion time $t$ through $\sigma=\sigma_{v}t$. %This construction
%assumes that all interaction energy has been converted into the kinetic energy  with no significant atom loss during expansion, and it provides a useful qualitative description of our data.
A blue OFR laser detuning near the -24\,MHz PA line \cite{zbl06}
increases $a$, leading to more interaction energy and larger expansion velocity and BEC size. Red detuning produces the opposite behavior. When the the total energy becomes negative, this simple explanation breaks down, and one observes  condensate collapse and significant loss of condensate atoms.

\begin{figure}
\includegraphics[clip=true,keepaspectratio=true,width=3.4in,trim=0.35in 0.1in 0.5in 0.3in]{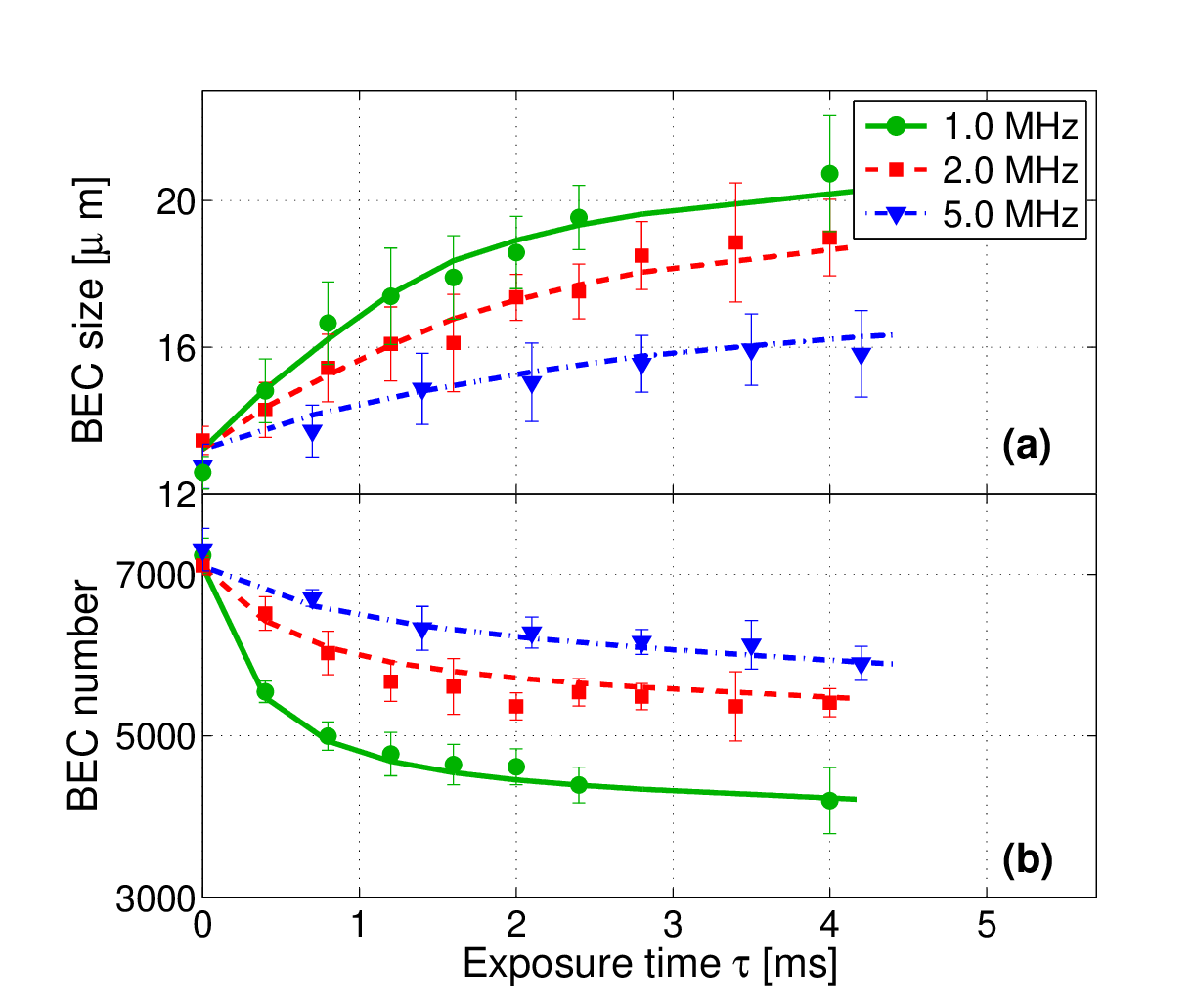}\\
\caption{(color online) (a) BEC size after 35 ms of expansion versus the exposure time of the OFR laser with the intensity of 0.057\,W/cm$^2$ and three different detunings from the -24\,MHz PA line. (b) Number of condensate atoms versus exposure time. Curves calculated by the Gross-Pitaevskii equation correspond to a combined fit of the data, yielding $\eta=19.5$, $\ell_{{\textrm{opt}}}/I=2.2 \times 10^4\,a_0/$(W/cm$^2$), and $K_0 = 5.8 \times 10^{-7}$\,cm$^3$/s. Error bars represent the standard deviation of the mean from multiple measurements.
 \label{BECsizeVsExposureTime24MHz} }
\end{figure}

In Fig.\,\ref{BECsizeVsExposureTime24MHz}, we study the variation of the BEC
size and number with the exposure time, $\tau$, for several blue detunings of the OFR laser. We observe that  several ms
is required for full conversion of the interaction energy into kinetic, with larger
detuning and smaller optically induced scattering length requiring longer $\tau$.
We can estimate the timescale for conversion with a hydrodynamic description of the condensate dynamics \cite{dgp99}.
The acceleration of atoms during expansion arises from the interaction pressure $P = gn(r)^2/2$, and
a characteristic acceleration $\tilde{a}$ can be approximated from $mn(r)\tilde{a} \approx -\nabla P \approx -n(r)\nabla[gn(r)]$.
This yields $\tilde{a} = -{\nabla[gn(r)]}/{m} \sim {g n_0}/{m \sigma_0}$.
In the large $N_0 a/a_{\textrm{ho}}$ limit with $a_{\textrm{ho}}=[\hbar/(m\overline{\omega})]^{1/2}$ , one can neglect the kinetic-energy term in Eq.\,\ref{OFR simple model} to find the characteristic final velocity given by the conservation of energy, $v_f \sim \sigma_v\sim \sqrt{{g n_0}/{m}}$. This implies a conversion timescale, ${v_f}/{\tilde{a}} \sim \sigma_0 \sqrt{{m}/{(gn_0)}}$,
%\begin{equation}\label{Conversion timescale}
%\tau = \frac{v_f}{\tilde{a}} \sim \sigma_0 \sqrt{\frac{m}{gn_0}},
%\end{equation}
 of 1\,ms for $a_{\textrm{opt}}$ of   $10\,a_0$, which roughly matches observations.
 Losses from single-atom light scattering preclude leaving the OFR beam on during the entire expansion time, and knowledge of the time required for close to full conversion is helpful for interpreting the results of experiments in which we apply the OFR laser for a fixed interaction time and vary the detuning, which will be discussed below.

To quantitatively analyze the variation of size and atom number versus interaction time and extract OFR parameters, it is necessary to treat dynamics and atom loss with the time-dependent non-linear Gross-Pitaevskii equation,
including the effects of $a_{{\textrm{opt}}}$, $K_{\textrm{total}}$, and single atom
light scattering, and neglecting effects of thermal atoms. The fit parameters are $\ell_{{\textrm{opt}}}/I$, $\eta$,  and $K_0$. %(Eq.\ \ref{eq:backgroundloss}).
The
rate of atomic light scattering  varies from 12 to 17\,s$^{-1}$, and is included in
the simulation assuming every scattering event results in the loss of one atom.

The fits are shown in Fig.\,\ref{BECsizeVsExposureTime24MHz}. The data at largest detuning from photoassociative resonance strongly determine the background loss because
loss from the OFR is small there.
%\textbf{($K_{in}< K_b=(2.0\pm xx)\times10^{-13}$ for $\Delta=5.0$\,MHz)}.
The fit optical length is
 $\ell_{{\textrm{opt}}}/I=(2.2\pm 1.0)\times 10^4 \,a_0/$(W/cm$^2$), and the fit parameter $K_0=(5.8\pm1.3) \times 10^{-7}$\,cm$^3$/s. Loss from the OFR is described by $\ell_{\textrm{opt}}$  and $\eta=19.5^{+8}_{-3}$, and there is strong anti-correlation between $\ell_{\textrm{opt}}$  and $\eta$. The uncertainty is dominated by systematic uncertainty in the trap oscillation frequency and imaging resolution.
These results are in good agreement with
the measured value  $\ell_{{\textrm{opt}}}/I=1.58\times10^4\,a_0/$(W/cm$^2$) and disagree slightly with $\ell_{{\textrm{opt}}}/I=8.3\times 10^3\,a_0/$(W/cm$^2$) calculated
directly from knowledge of the molecular potentials \cite{bnb11}.

Experiments with a thermal strontium gas \cite{bnb11} found larger losses associated with an OFR than described by theory, which was described by $\eta=2.7$. These measurements probed  the core of the photoassociative transition ($|\Delta|<50\,\Gamma_{\textrm{mol}}$). The additional loss is not well understood. We see a similar resonance width in a BEC when we significantly reduce the laser intensity and interaction time and take a photoassociative loss spectrum  of this core region.  Our use of the  OFR probes the distant wings ($50\,\Gamma_{\textrm{mol}}\,<\,\Delta\,<\,667\,\Gamma_{\textrm{mol}}$), and a fit of the loss using the single resonance model requires an even larger value of $\eta$.
%Perhaps this large value reflects that fact that the loss is transitioning from being dominated by the core $K_{in}$ to being dominated by the background $K_b$. Alternatively, one could
We interpret the varying $\eta$ values as meaning that the  full spectrum of photoassociative loss, including the far wings, is not well described by a Lorentzian.

\begin{figure}
\includegraphics[clip=true,keepaspectratio=true,width=3.65in,trim=0.4in 0in 0in 0.in]{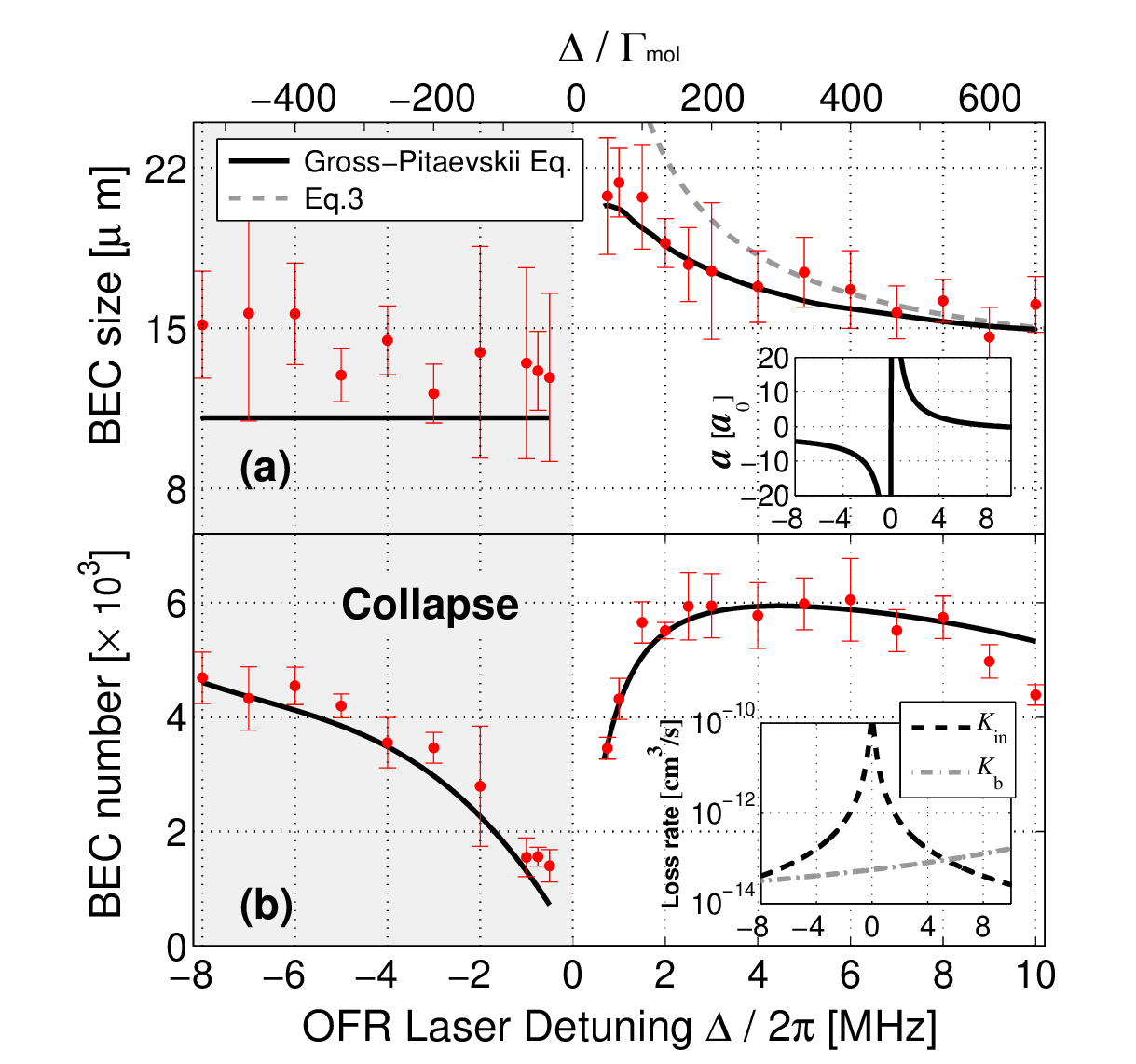}\\
\caption{(color online) The BEC size(a) and number(b) versus the detuning with respect to the -24 MHz PA resonance for an intensity of 0.057\,W/cm$^2$. The OFR beam is applied for 4.0\,ms, and the data are recorded after 35 ms of expansion. The insets give the total scattering length $a$ and the loss rate constants.
\label{BECsizeNumVsOFRdetuning} }
\end{figure}

The dependence of the BEC size and number on detuning from the -24\,MHz PA line is shown in Fig.\,\ref{BECsizeNumVsOFRdetuning} for a fixed intensity and interaction time $\tau=4$\,ms. The fit parameters from Fig.\,\ref{BECsizeVsExposureTime24MHz} describe the data well over this range. Note that the number of atoms initially increases with blue detuning from PA resonance as the loss from the OFR ($K_{\textrm{in}}$) decreases. The number then slowly decreases because the background loss ($K_\textrm{b}$) increases approaching atomic resonance.
The BEC size data predicted
by  Eq.\,\ref{OFR simple model}, which neglects atom loss and assumes that
the OFR laser is applied long enough to fully convert interaction energy into kinetic, is also
shown in Fig.\,\ref{BECsizeNumVsOFRdetuning}a. The difference between this curve and the data highlights that
 atom loss is significant during the conversion process at smaller detunings, and
the Gross-Pitaevskii equation simulation  is required to describe the data. A typical total scattering length (Fig.\,\ref{BECsizeNumVsOFRdetuning}a inset) is $a=20\,a_0$ for $\Delta=2\pi\times1$\,MHz\,$\simeq\,67$\,$\Gamma_{\textrm{mol}}$.

For red detuning, the OFR laser makes the scattering length more negative and triggers a collapse of the
condensate, which is evident as large loss
in the plot of condensate number remaining after expansion (Fig.\,\ref{BECsizeNumVsOFRdetuning}b). The dramatic asymmetry
of loss with respect to detuning from resonance shows that the loss must reflect condensate dynamics \cite{kms98,gsp00,dcc01}, not photoassociative loss directly caused by the OFR laser.
The Gross-Pitaevskii equation provides a good description of the BEC number data for red detuning in spite of the fact that the collapse dynamics may contain beyond-mean-field effects \cite{dew96} not taken into account in the Gross-Pitaevskii formalism.

%The Gross-Pitaevskii equation provides a surprisingly good description of the  number of atoms remaining for red detuning. Condensate collapse introduces additional loss processes \cite{kms98} and one expects  beyond-mean-field effects \cite{dew96} in this regime that are not treated by the Gross-Pitaevskii formalism.

A variational calculation of the condensate energy functional as a function of
condensate size  \cite{pmc97,dgp99} for the parameters of Fig.\,\ref{BECsizeNumVsOFRdetuning} predicts that the condensate  expands
initially after the trap is extinguished if $a > -3.8\pm
0.2$\,$a_0$. For more negative $a$ ($-10\pm3$\,MHz\,$<\Delta/2\pi<0$\,MHz), there is no
repulsive energy barrier on the effective potential for the system and collapse
results.
%This would yield collapse for detuning $-10\pm3$\,MHz\,$<\Delta/2\pi<0$\,MHz,
%where the uncertainty results from uncertainties in the trap potential and
%$\ell_{\textrm{opt}}/I$ considered in Fig.\,\ref{BECsizeNumVsOFRdetuning}, which is in
%rough agreement with our observations.
Numerical simulation of the
Gross-Pitaevskii equation supports this interpretation. Simulations show that collapse can be very non-uniform, as predicted in
\cite{kms98}, with significant density increase only near the condensate center  for $a$ only moderately more negative than the
threshold. %This may explain the continuous reduction in loss as red detuning is
%increased. It also implies that it is difficult to rigorously state a
%threshold for collapse of the released condensate.

In summary, we have demonstrated control of collapse and expansion of an $^{88}$Sr BEC using an intercombination-transition OFR. At large detuning from PA resonance ($\lesssim667,\Gamma_{\textrm{mol}}$), we obtain sample lifetimes on the order of 1 ms while changing the scattering length by 10's of $a_0$.
 While this is a moderate change compared to the mean scattering length \cite{gfl93} for Sr,
 $\bar{a}=[4\pi /\Gamma(1/4)^2](1/2)(2\mu C_6/\hbar^2)^{1/4}=75.06$\,$a_0$, it is an extremely large relative change for $^{88}$Sr ($a_{\textrm{opt}}/a_{{\textrm{bg}}}=\pm10$) because of the small $a_{\textrm{bg}}$. The OFR can thus drastically change the dynamics. Here,
$\Gamma(x)$ is the gamma function, and $C_6=3170$\,a.u. is the van der Waals coefficient for the interaction between two ground state Sr atoms \cite{pde02} in atomic units.

Our work probes collisions of atoms in a light field in a previously unexplored region of large detuning from photoassociative resonance. The isolated resonance model
\cite{ctj05,ctj06} provides a good description of the optically induced scattering length (Eq.\,\ref{OFRFormulas}) out to a detuning of $|\Delta|\,\simeq\,667\,\Gamma_{\textrm{mol}}$ for this photoassociative transition. This is not surprising
because the detuning from the PA resonance
is still much less than the spacing between excited molecular states.
A coupled channels numerical calculation \cite{bnb11} shows the breakdown of the isolated resonance approximation and absence of a significant OFR
effect at comparable detuning from two PA lines.
%We also observe this
%behavior. In particular, we observe no OFR effect for blue detuning from the
%least bound PA line, which is very close to atomic resonance.  According to the isolated resonance model, this PA transition would be expected to produce a large
%effect due to its extremely large Franck-Condon factor \cite{zbl06}.
The isolated resonance model is valid over a much smaller range for describing the loss induced by the OFR laser because of the background loss and the enhanced loss parameterized by a large value of $\eta$ in the far wings of the line.

The original peak density of the condensate is extremely high in our experiment because of the attractive interactions. Increased lifetime or larger OFR effect should be obtainable for densities commensurate with single-site loading of an optical lattice. Improvements could also be made by working at larger detuning from PA resonance and larger laser intensities. Working with a more deeply bound excited molecular state such as the PA line at $-1.08$\,GHz \cite{zbl06} may offer advantages in this direction, such as greater suppression of atomic light scattering and reduced background two-body loss. This holds promise to bring many possible experiments involving optical Feshbach resonances and quantum fluids into reach.

We thank Paul Julienne for helpful discussions and acknowledge support from the Welch Foundation (C-1579 and C-1669) and the National Science Foundation (PHY-1205946 and PHY-1205973).

%\bibliography{bibliography,library}

\end{document}